\documentclass[]{spieman}
\usepackage{graphicx}
\usepackage{amsmath}
\usepackage[titletoc]{appendix}
\usepackage[]{graphicx}
\usepackage{setspace}
\usepackage{tocloft}

\title{CCD detectors with high QE at UV wavelengths}

\author{Erika~T.~Hamden\supscr{1}, April~D.~Jewell\supscr{2}, Charles~A.~Shapiro\supscr{2}, Samuel~R.~Cheng\supscr{2}, Tim~M.~Goodsall\supscr{2}, John~Hennessy\supscr{2}, Michael~Hoenk\supscr{2}, Todd~Jones\supscr{2}, Sam~Gordon\supscr{3}, Hwei~Ru~Ong\supscr{4}, David~Schiminovich\supscr{4}, D.~Christopher~Martin\supscr{1}, Shouleh~Nikzad\supscr{2}}

\affiliation{\supscrsm{1}Department of Astronomy, California Institute of Technology, 1200 East California Blvd, Pasadena, CA 91125, USA \\ 
\supscrsm{2}Jet Propulsion Laboratory, California Institute of Technology, M/S 302-304, Pasadena, CA 91109, USA \\
\supscrsm{3}Arizona State University, School of Earth and Space Exploration, Tempe, AZ, 85287 USA\\
\supscrsm{4}Department of Astronomy, Columbia University, 550 W 120th St, New York, NY 10025, USA}

\cftpagenumbersoff{figure}
\cftpagenumbersoff{table} 
\begin{document}
\maketitle

\begin{abstract}

We report on multi-layer high efficiency anti-reflection coating (ARC) design and development for use at UV wavelengths on CCDs and other Si-based detectors. We have previously demonstrated a set of single-layer coatings which achieve $>$ 50\% quantum efficiency (QE) in 4 bands from 130 to 300 nm. We now present multi-layer coating designs that significantly outperform our previous work between 195 to 215 nm. Using up to 11 layers, we present several model designs to reach QE above 80\%. We also demonstrate the successful performance of 5 and 11 layer ARCs on silicon and fused silica substrates. Finally, we present a 5 layer coating deposited onto a thinned, delta-doped CCD and demonstrate external QE greater than 60\% between 202 and 208 nm, with a peak of 67.6\% at 206 nm.
\end{abstract}

{\noindent \footnotesize{\bf Address all correspondence to}: Erika T. Hamden, California Institute of Technology, 1200 East California Blvd, MC 278-17, Pasadena, CA, USA, 91125; Tel: +1 626-395-5711; E-mail: \linkable{hamden@caltech.edu}}

\begin{spacing}{2} 
\section{Introduction}

Since their invention at Bell Labs in 1969, charged coupled devices (CCDs) have become ubiquitous in astronomy. These silicon-based imagers have had great success in both ground and space-based applications at wavelengths as varied as the X-ray, visible, and near IR. Until recently, CCDs have not been effective in the UV (100-350 nm), with low to zero quantum efficiency (QE). This low efficiency is due to both the very shallow absorption depth of UV photons and the absorptive nature of the poly-silicon front side circuitry. A combination of thinning, backside illumination, and a process called delta doping can bring the QE of CCDs up to the reflection limit of silicon at all wavelengths in a stable and consistent way \cite{1992Hoenk,1994Nikzad,2013Hoenk}; in the UV this brings the QE to around 30\% between 150 and 300 nm. 

NASA's Jet Propulsion Laboratory (JPL) has had great success in the development of delta-doping technology. This technique passivates the back surface of a thinned silicon CCD, complementary metal-oxide semiconductor (CMOS) imagers, or p-type/intrinsic/n-type (PIN) diode array and results in nearly 100\% internal QE from the extreme UV to the IR \cite{1992Hoenk}. The shallow absorption depth of UV photons in silicon ($< 10$ nm) requires the use of both back-illuminated and thinned CCDs, to avoid both absorption in the front-side circuitry and recombination loses in the backside surface layer. Without some type of passivation, such as delta doping, traps will form in the surface layer between Si-SiO$_2$ and interact with the shallowly generated charges. This interaction causes poor QE, hysteresis, and high dark current \cite{1989Janesick}. The insertion of a single atomic layer of heavily doped silicon, deposited using molecular beam epitaxy (MBE) and capped with pure silicon, passivates the surface layer. This `delta-doped' layer eliminates traps and allows for stable and efficient charge collection within the CCD \cite{1994Nikzad}. Delta doping provides a superior passivation to other methods that can be destructive to the silicon crystal or have only short-lived effects. While delta doping modifies the internal electric field of the device, the surface itself remains as silicon covered by its native oxide. The QE of a delta-doped device is limited by reflection from the silicon surface. Losses due to reflection can be reduced with the application of a suitable anti-reflection coating (ARC).

Single layer ARCs are often used to increase transmission over wide band passes in the visible (e.g. HfO$_2$ for increased blue sensitivity \cite{1987Lesser}). We have previously shown $>$50\% QE at UV wavelengths (100-300 nm) using single layer AR coated delta-doped detectors \cite{2011Hamden,2012Nikzad}, an unprecedented QE performance for any UV detector, especially between 100 and 250 nm. These high-quality ARCs were made possible with atomic layer deposition (ALD, see Ref.~\citenum{2009George} for an excellent introduction to ALD and it's applications). Work by Ref.~\citenum{2007Westhoff} using MBE to passivate the backsurface of a thinned CCD has yielded high QE ($\sim$50\% at 100 nm), but testing focused primiarly on the extreme UV ($<$ 110 nm) and did not include ARCs. More standard, commercially available blue enhancing coatings on backside illuminated CCDs are able to achieve high QEs above 250 nm. 

For higher transmission, quarter wave stacks and other configurations will minimize reflectance to nearly zero over a limited range of wavelengths and have been well explored in the visible \cite{1986Macleod}. These multi-layer techniques can be extended to UV wavelengths, but are constrained by both the limited range of UV-transmissive materials and by the changing index of refraction of silicon (varying from 0.5 to 5.5 between 100 nm and 350 nm). Silicon itself is an absorbing substrate with a high complex refractive index and, therefore, the ARC requirements are different than for a more conventional transmissive coating. Further complicating matters are the trade-offs between achieving a high QE through the use of multi-layers, narrowing of the band pass, and the potential for increased absorption in the UV as the overall coating thickness grows. 

For this work, we focus on designing multi-layer coatings for use in the UV and optimized over a narrow wavelength range (195-215 nm). This 20 nm band is advantageous from a materials standpoint, as the index of refraction of silicon is relatively unchanging over this narrow range. Furthermore, the nature of the multi-layer ARC makes wider band passes more difficult to design. A number of materials can be deposited via ALD with limited absorption in this band pass, providing several options for good ARC designs. Most importantly, this wavelength range offers particular advantages since it corresponds to an atmospheric window between O$_2$ and O$_3$ absorption bands \cite{1965Brewer}, permitting observations in the UV at balloon and rocket altitudes. The ideal test bed for a thinned, delta-doped and AR coated CCD is the Faint Intergalactic Redshifted Experiment Balloon (FIREBall, Ref.~\citenum{2008Tuttle}) experiment, set to launch in 2016, which will measure emission in the 200-210 nm range and benefits from high QE, UV-optimized ARC.

The rest of this paper is organized as follows: in Section \ref{sec:Models}, we describe techniques for model creation and the model selected for deposition. Section \ref{sec:Deposition} details deposition recipes and techniques, along with growth rates for each material. Section \ref{sec:results} presents the reflectance and transmission measurements of the coatings on inert substrates. Section \ref{sec:CCD} covers deposition, testing and results of a coating deposited on a delta-doped CCD. In Section \ref{sec:discussion}, we discuss the implications of our results and prospects for future work.

\section{Model Development}
\label{sec:Models}

We have developed a suite of possible coatings designed to maximize QE within the FIREBall band. ARC performance was modeled using TFcalc\textregistered (Software Spectra, Inc.); materials under initial consideration included MgO, MgF$_2$, Al$_2$O$_3$, and SiO$_2$. HfO$_2$, often used for visible ARCs, was rejected due strong absorption below 220 nm. ARC performance was modeled based on optical constants from Ref.~\citenum{1985P,1991P}, as well as vacuum ellipsometry measurements performed in house by J. A. Woollam (VUV-VASE) using samples prepared at JPL. The substrate was modeled using the measured indices for delta-doped silicon, which differs slightly from that of un-doped silicon (see Appendix). Figure \ref{fig:sidd} shows the difference in reflection between bulk silicon and delta-doped silicon; while the difference is slight at most wavelengths, the reflectance can change by several percent at UV wavelengths. Our models also included a thin (~2 nm) native oxide at the substrate interface. 

Our ARC designs are summarized in Table \ref{tab:char} and plotted versus wavelength in Figure \ref{fig:altlayers}; we present peak QE, the corresponding wavelength, materials used, and the width of band at 50\% QE. For simplicity, initial ARC designs were restricted to three layers and only two materials; this provided a basis for further optimization with additional layer pairs. Four of the six starter designs yielded potential transmission above 80\% at 205 nm. Of these, 3C was selected as the basis of multi-layer coatings with more than three layers, using SiO$_2$ and Al$_2$O$_3$ in large part because their growth is straightforward and repeatable using ALD. Table \ref{tab:char} also includes the details for two more complex coatings, using five (5A) and eleven (11A) layers. 5A is similar to 3C, but with a higher peak and narrower width. 11A has two close peaks with lower transmittance than the single peak of coating 5A, but higher transmittance at the edges of the band (above 70\% between 195-215 nm), as shown in Figure \ref{fig:11layer}. 

More complex ARC designs with three or more materials may offer higher QE or broader peaks than those discussed in this paper; however, the feasibility of more complex models is limited by physical constraints. For example, Al$_2$O$_3$ begins absorbing at $\sim$240 nm, depending on film preparation, and is strongly absorbing below 180 nm; this will impact transmission as the number and thickness of the Al$_2$O$_3$ layers increase. Other materials, such as HfO$_2$, MgO, and TiO$_2$ have well known absorption cutoffs near UV wavelengths, and so are not as useful over a large wavelength range. Even so, we show here that the transmittance peak can be shifted to nearly any wavelength above the absorption cutoff by modifying the film thicknesses or the materials used. Further discussion of this cutoff for a range of UV film materials can be found in Ref.~\citenum{2011Hamden}.

\section{Deposition Techniques}
\label{sec:Deposition}

ALD enables conformal, atomic level control over coating growth and allows well-controlled, repeatable depositions. Spatial uniformity of ALD films can be affected by a variety of parameters, including system pressure, pulse and purge times, substrate temperature and surface preparation (i.e., cleanliness). For example, incomplete purging or non-uniform heating may lead to parasitic CVD-like growth, resulting in non-uniform coatings. The growth recipes used in in this work showed repeatable “ALD” growth behavior, and film quality was verified prior to deposition on live devices. Initial ALD recipes for Al$_2$O$_3$ came from Ref.~\citenum{2008Goldstein}, MgO from a custom reaction developed at JPL by the authors, SiO$_2$ from Ref.~\citenum{2011Dingemans} and HfO$_2$ from Ref.~\citenum{2005Liu}. Growth recipes required modifications to accommodate the equipment used. Film composition and uniformity were validated using X-ray photo-electron spectroscopy (XPS), transmission electron microscopy (TEM), secondary ion mass spectrometry (SIMS), spectroscopic ellipsometry, or a combination of these techniques. Growths were conducted at $200\,^{\circ}\mathrm{C}$. Additional details regarding the ALD growth parameters can be found in Ref.~\citenum{2013Greer}. A Beneq TFS200 ALD system at JPL was used for all depositions. Our work with ALD spans several years and systems. 

Coatings were deposited on 1 inch $<$100$>$ 1-20 ohm-cm silicon wafers (to test reflectance) and on 3-mm thick fused silica window with $\lambda$/10 peak-valley flatness at 635 nm (to test transmission/absorption). For simplicity, all measurements (and corresponding models) plotted in the following figures are based on a silicon substrate without the doped layer; previous work on delta-doped CCDs has shown good fidelity between both model predictions and results from silicon wafer substrates \cite{2012Nikzad,2011Hamden}. Film thicknesses were verified by spectroscopic ellipsometry (using either a Sentech SE850 or Horiba UVISEL 2).

Our previous work measuring QE, described in Ref.~\citenum{2012Nikzad} and Ref.~\citenum{2011Hamden}, consisted almost entirely of single-layer coating deposits on silicon substrates. In general, ALD deposition rate calibration is done by measuring the thickness of a single layer on a silicon substrate after a certain number of cycles. However, ALD film nucleation and growth rates for a given material will often vary depending on the substrate. For the multi-layer coatings discussed in this paper, only the first layer is deposited directly on silicon; subsequent layers were deposited onto other ALD films. As a result, we found that the SiO$_2$ layers within the stack were consistently 3-4 nm thinner than desired. Additional experiments revealed that the nucleation rate of SiO$_2$ on Al$_2$O$_3$ is different than that on Si- that is the number of cycles required to achieve good surface nucleation is greater. Figure \ref{fig:nucleation} shows the measured thickness for growths of ALD-SiO$_2$ on bare silicon and on silicon with a 7 nm layer of Al$_2$O$_3$. In all cases, the SiO$_2$ grown on the Al$_2$O$_3$ base was thinner for the same number of cycles. The nucleation delay can be seen as an offset in the y-intercept. With good surface nucleation, the deposition rates were nearly identical (1.12 vs 1.15 \AA\ per cycle). In practice, several additional cycles were required when growing SiO$_2$ on Al$_2$O$_3$ versus silicon or its native oxide. We found no measurable difference in the indices of refraction (n and k) for SiO$_2$ grown on Al$_2$O$_3$ versus silicon, nor for AL$_2$O$_3$ grown on SiO$_2$ versus silicon.

\section{Reflectance and Transmittance Testing}
\label{sec:results}

Before applying these ARCs to functional CCDs, we first tested them out on inert substrates for performance verification. This provides a consistency check for coating performance prior to the more complicated direct QE measurements. We wanted to verify that the coating alone performed as expected without the complicating factors of the CCD operation. Once these optical tests matched the model performance, then we applied several ARCs to functional CCDs. Any major discrepancies between the QE measurements and the model are likely the result of device operation and not the coating performance. 

Reflectance and transmittance tests were conducted with the same system at Columbia University. The samples were placed in a vacuum chamber maintained at less than 1 $\times$ 10$^{-4}$ Torr for the duration of the measurement. An Acton Research Corp VM-504 monochromator, fed by a focused deuterium lamp, provides illumination to the samples. The light is reflected from silicon substrates for measuring reflectance or is directed through transparent substrates for measuring transmission. 

Reflectance measurements were performed at $5-10\,^{\circ}$ from normal incidence with silicon substrates. We recorded intensity using a PMT (R6095) with a scintillator and light pipe assembly (McPherson Model 658). For each sample, the following set of measurements were made: direct intensity from the lamp, reflected intensity from the sample, reflected intensity from a bare silicon standard, reflected ambient intensity not directly in the path of the light. This last measurement is from an open filter position and serves as a background measurement. These measurements were then used to calculate direct reflectance from the sample and silicon (as a standard).

Transmission measurements were performed at normal incidence with fused silica substrates. With the same set-up as for reflectance measurements, we recorded the following measurements: direct intensity from the lamp, reflected intensity from the sample, direct transmittance through the sample, reflected intensity from a bare fused silica window, direct transmittance through a bare fused silica window, and reflected ambient intensity as a background measurement. We also re-modeled our selected ARCs with fused silica substrates to provide models of expected transmission and reflectance, since this will differ significantly from models with a silicon substrate.

Transmission into the Si is calculated by measuring both reflectance from the ARC with a silicon substrate and verifying reflectance and transmission from the same ARC applied to a fused silica window. While fused silica is an incoherent substrate with a non-negligible thickness, we felt that absorption/transmission measurements on an alternative substrate could provide a more nuanced view of the coating performance, despite not being a perfect match to the silicon substrate. The models used to verify the performance of the coating on fused silica take into account the thickness of the window and use the expected indices of refraction for the individual coating layers. Measurements of reflectance and transmission through the ARC on fused silica match the model with good fidelity (see bottom panels of Figures \ref{fig:5layer} and \ref{fig:11layer_data}) and confirm that the behavior of the ARC on a different substrate is as expected in both transmittance and reflectance. 

To provide a better idea of expected transmission based on our testing, we use the fused silica substrate measurements to provide an idea of absorption attributable to the coating itself. This process is in lieu of measuring absorption directly, which can only be done via QE testing. To estimate absorption in the layer, we make several assumptions. First, we assume that the absorption of a multi-layer ARC on Si will be nearly the same as the ARC on fused silica, and thus we can use one to approximate the other. The models show this to be a reasonable assumption. We also assume that there is not significant absorption in the ARC at our wavelength of interest and above (absorption should be less than 10\% above 195 nm). We also treat the coating as a single body to simplify the calculation. Since potential transmittance is multiplicative through several layers \cite{1986Macleod}, we assume that the potential transmittance of the ARC on fused silica is equal to the potential transmittance of the ARC times the potential transmittance of the fused silica, where potential transmittance is defined as:

\begin{equation}\label{eq:transeq}
 \psi_j = \frac{T_j} {(1-R_j)} \\
\end{equation}

for a layer j. Thus to determine the absorption of the ARC alone, we divide the measured absorption of the layer and window by the measured absorption of only the window. This does not well approximate the absorption of the layer at very low wavelengths ($<$ 160 nm), where the fused silica is very absorbing. However, since our region of interest is at 195 nm and longer, we don't expect this will significantly change our results. For shorter wavelength targets, a MgF$_2$ window can be used. Estimated transmission is calculated by subtracting this measured absorption and previously measured reflectance of a coated silicon substrate from 100\%, as shown in Equations \ref{eq:transcal1} and \ref{eq:transcal2}.

\begin{equation}\label{eq:transcal1}
 (1-A_{coating}) = \frac{(1-A_{coating,fused Si})} {(1-A_{fused Si})} \\
\end{equation}
\begin{equation}\label{eq:transcal2}
 T_{calc} = 1.0 - R_{Si,coating} - A_{coating} \, ,
\end{equation}

where $A_{coating}$ is calculated absorption due to the multi-layer coating only, $A_{coating,fused Si}$ is measured absorption due to both the coating and the fused silica window, $A_{fused Si}$ is measured absorption of an uncoated fused silica window, $T_{calc}$ is the calculated transmission of the overall coating, and $R_{Si}$ is the measured reflection off the coated silicon substrate. For small absorbances in the ARC, (i.e. low values for A$_{coating}$), Equation \ref{eq:transcal1} is valid given the assumptions described above and provides a good fit when compared to the expected transmission.

In this paper we preset results from two coating designs, 5A and 11A. For coating 5A, the overall performance matched the model with high fidelity, as shown in Figure \ref{fig:5layer}. The minimum reflectance point originally expected at 205nm was shifted towards lower wavelengths, due to the SiO$_2$ growth discrepancy described in Section \ref{sec:Deposition}. We modified the thickness of the components in the initial model to create a version that best matches the measured coating performance. We are able to model the new coating with thinner SiO$_2$ layers, and find that it matches the observed reflectance and expected transmission.

To further study the effect of increased multi-layer stacks, we also tested an 11-layer coating (model 11A), which is designed to have a double peak, as shown in the model (Figure \ref{fig:11layer}) and measurements (Figure \ref{fig:11layer_data}). This coating was prepared incorporating the rates of SiO$_2$ growth on an Al$_2$O$_3$ base layer. ARC design 11A has a slightly lower theoretical peak QE (84\% vs. 90\%) but has higher average transmission in the 20 nm band pass. The estimated transmission is somewhat off from the model, with absorption calculated as described above for the 5-layer coating. This slight change is likely the result of a few nanometers difference in the top or bottom layer thickness. Changes of only a nanometer or two can be enough to change the peak heights by several percentage points.

\section{CCD ARC deposition and QE measurements}
\label{sec:CCD}

After our initial tests on silicon substrates, we applied a variation of coating 5A (accounting for the SiO$_2$ growth rate on Al$_2$O$_3$ as discussed earlier), described in Section \ref{sec:results}, to a thinned and delta-doped e2v CCD201-20s. The preparation and delta-doping process are described in more detail previously \cite{2012Nikzad,2013Hoenk,2014Hoenk}. For our work, devices were processed at wafer level, but the same processing can be performed on individual die. Wafers are first bonded to a carrier wafer to provide support after thinning. This bonding process is performed by Ziptronix. Bonded wafers are then thinned to the epitaxial layer using a custom process developed at JPL. After thinning, the wafers are cleaned and prepared for delta-doping. After delta-doping the wafer can either be diced into individual devices for custom ARC deposition or a single ARC can be deposited on the entire wafer and then diced. In either case, individual devices are then cleaned, inspected, and packaged for testing. Operation of these devices follows the standards recommended by the manufacturer.

The characterization set-up has also been previously described in detail \cite{2011Jacquot}. To briefly summarize, this set-up uses a careful calibration system to directly measure QE from a range of devices. A deuterium lamp is used for wavelengths below 350 nm and a quartz tungsten halogen lamp above 350 nm and into the IR. A set of long pass and band pass filters minimize leakage from bluer wavelengths in the case of the deuterium lamp, and redder wavelengths for the quartz tungsten lamp. These filters are sufficient to reduce out of band light to less than 2 percent, and below 0.5 percent at UV wavelengths. The filtered light passes through a vacuum monochromator that has two output slits selectable by a flip mirror. One slit feeds either the calibrated photodiode (PD) or a cooled CCD for QE testing. The other slit feeds a second "concurrent photodiode" that takes flux measurements interleaved in time to account for lamp variability.  The entire slit image is captured by each PD or CCD without overfilling.

To measure QE, the CCD and a fast shutter are used to image the slit at each wavelength, and calibration dark frames are taken with the same exposure times as the slit images.  Interleaved photocurrent measurements $I(\lambda)$ are taken with the concurrent PD.  The calibrated PD is used to measure photocurrent at the same wavelengths, also interleaved with the concurrent PD.  We define $R(\lambda)$ as the ratio of the two PD measurements, which only depends on the throughput ratio of the 2 output slits and the QE ratio of the 2 PDs.  CCD conversion gain $K$ (e$^-$/Digital Number, DN) is calculated using the standard photon-transfer curve procedure \cite{2007Janesick}.  Equation \ref{eq:QEcalc} summarizes the QE calculation, defined as the total CCD signal in e$^-$ divided by the expected number of photons in the exposure:

\begin{equation}
\label{eq:QEcalc}
\begin{aligned}
QE_{CCD}(\lambda) = \frac{S(\lambda) QE_{PD}(\lambda)}{I(\lambda) R(\lambda)} \frac{KC}{\Delta t} \, ,
\end{aligned}
\end{equation}

where $S(\lambda)$ is the integrated signal on the CCD (DN), QE$_{PD}(\lambda)$ is the QE of the calibrated PD, $C$ is the elementary charge in Coulombs, and $\Delta t$ is the exposure time of the CCD image. Knowing QE$_{CCD}$, an observer using the CCD in the field can estimate a source's photon flux from the CCD signal in e$^-$.

At wavelengths below 400 nm, photo-charges can be produced with enough kinetic energy to send additional e$^-$ into the conduction band through collisions. While the QE in Equation \ref{eq:QEcalc} above correctly relates the detector signal to the incident photon flux, obtaining the fraction of incident photons detected requires a correction to account for the generation of multiple electron/hole pairs per photon \cite{2012Nikzad}. The average number of electron/hole pairs generated per interacting photon is known as quantum yield (QY), and random variation around the average is known as Fano noise \cite{1988Janesick,2007Janesick,2008McCullough}. The equation for correcting for QY is as follows:

\begin{equation}
\label{eq:QYcalc}
\begin{aligned}
QE_{corrected}(\lambda) = \frac{QE_{CCD}(\lambda)}{QY(\lambda)} \, ,
\end{aligned}
\end{equation}

where QE$_{CCD}$ is the QE as calculated using Equation \ref{eq:QEcalc}, QY is the quantum yield, and QE$_{corrected}$ is the true measure of the fraction of interacting photons.

Measurements of QY in silicon use one of two methods. The first method, used by Kuschnerus, is to compare the detector QE to the measured or modeled transmission, assuming that QE accounts for any discrepancy between them \cite{1998Kuschnerus,1998Canfield}. This method provides an upper limit of QY, since it relies on the assumption that the photodiodes have small internal losses and that the silicon-oxide interface has limited absorption. The assumption regarding absorption of the oxide fails below 160 nm, but these results provide reasonable QY for longer wavelengths. An alternative method \cite{2007Janesick,2010Borders} uses a photon transfer curve (PTC) taken at relevant wavelengths. This method exploits the fact that the mean and variance of the detector signal scale as QY and QY$^2$ respectively, violating the behavior of the photon shot noise, which is Poisson-distributed. PTC measurements of QE are accurate at very high energies (e.g. X-ray) where the e/h pairs generated per photon are strongly peaked around a mean value, i.e. Fano noise is relatively low. In the UV, Fano noise contributes significantly to the signal variance, which biases the QY measurement (see equation 30 in McCullough\cite{2008McCullough}). Quantum yields calculated in this way are can be well below those calculated using the reflectance method (1.1 electrons/photon vs. 1.3 electrons/photon at 200 nm, for example) and can change the apparent QE of a detector significantly, as seen in Figure \ref{fig:QE}.

For the purposes of this paper, we use the values derived by Kuschnerus\cite{1998Kuschnerus} as a conservative measure of QE. Kuschnerus reports the mean energy W to produce an electron-hole pair in silicon in Figure 6 of their paper. We then estimate QY as:

\begin{equation}
\label{eq:QYKruschnerus}
\begin{aligned}
QY = \frac{E_{photon}}{W} = \frac{hc}{\lambda W}
\end{aligned}
\end{equation}

These QY values are higher than those measured using a PTC, and provide a lower limit of the possible QE measured. To provide an idea of the range in QE values, we also include QY values measured using the PTC method at JPL (green line in Figure \ref{fig:QE}), but we restrict our discussion to QE derived with the Kuschnerus values. A detailed study of QY in the UV, variations between detectors, and the exact contribution of Fano noise is needed to provide a more exact value for quantum yield and is beyond the scope of this paper.

The measured QE is compared to the expected QE based on the transmission model in Figure \ref{fig:QE}. The peak is located at 206 nm, with a maximum QE of 67.6\% at this wavelength. QE is above 50\% across the FIREBall band pass (200-210 nm) and above 60\% between 202 and 208 nm. Structure in the model, including the width of the peak at 205 nm and the increasing QE at longer wavelengths, are clearly present in the data. 

Previous tests of delta-doped devices have shown no deterioration with time \cite{1994Nikzad}. The coating and device presented here have been tested several times since the initial QE test presented here. We found no significant degradation in performance. In addition, QE tests performed at e2v on other devices showed no change in performance over an 8 month period. Our work with these devices has been limited to low or moderate intensity light. Devices prepared by our group for use in laser applications have worked well and are reportedly robust, but we have limited test data on their performance over time.

The deviation between the measured QE and the model could be the result of several processes. There is likely some absorption in the layers, which becomes more pronounced at UV wavelengths. Al$_2$O$_3$ begins to absorb significantly below 180 nm and there is some absorption in the FIREBall-2 band. Other factors, such as the presence of surface oxides or other contaminants during deposition, may account for some of the difference. Finally, we reiterate that there is significant uncertainty in QY measurements for Si in the UV. Because detector QE is a crucial component of relating observed detector counts to source flux, it's vital that QY measurements in the UV be improved if silicon-based imagers are to be useful in the UV.

\section{Discussion}
\label{sec:discussion}

Detectors with ARCs prepared by ALD exhibit QE that matches the overall shape of the model quite well. The exact QE is highly dependent on the quantum yield correction. Using a conservative QY correction, we measure a peak QE of 67.6\% for a 5-layer coating of SiO$_2$ and Al$_2$O$_3$ on a delta-doped electron multiplying CCD. The QE for this coating is above 50\% in the FIREBall band pass (200-210 nm), with an average QE of 61\% across the band. This result is a $>$25\% improvement in the measured QE over our previous work, in which we demonstrated 53\% QE at 205nm for delta-doped CCD with a single Al$_2$O$_3$ layer.

The basic structure of this design, five alternating layers of high- and low-index material, can be modified to fit nearly any wavelength. By changing the overall coating thickness and/or the materials themselves, the QE peak currently at 206 nm can be shifted to longer or shorter wavelengths. Below 150 nm, only metal fluorides, including MgF$_2$, LaF$_3$, and CaF$_2$ have low enough absorption to be practical for ARCs. 

In general, the addition of layers to an ARC provides the designer with increased flexibility. In the visible and IR wavelength ranges, added complexity can provide both improved transmission and broader peaks. In the UV, this is not necessarily true. In the case of the 3- and 5-layer coatings, the transmission peak width is narrowed significantly by the addition of two layers, dropping from 47 nm to 23 nm for 3A and 5A, respectively. For the FIREBall-2 balloon project, this narrow width matches the existing scientific requirement. Within the range of available materials, there is frequently some absorption in the UV which prevents one from achieving 100\% QE. Additionally, the indices of refraction of silicon are changing rapidly in the UV, but so are most of the coating material indices. The use of more than two materials may help expand the wavelength range beyond where good transmission ($>$ 60\%) can be reached with a single coating design, but we have not fully explored the limitations of broadband designs.

Without having a single broadband coating, high QE in the UV can still be achieved via a few different methods. The first is simply to restrict ourselves to a narrow band pass and achieve as high QE as possible, as we have done here. A second is to create band passes based on the indices of refraction of silicon. Several wavelength regions have relatively flat or smoothly varying indices and could provide natural band pass edges. 

A third method is to apply different coatings to different regions of a single detector. In the work described here, the entire CCD surface is uniformly AR-coated and only one detector is used, but this does not necessarily need to be the case. For a spectrograph, different wavelengths of light will be directed to particular regions of the detector. One could use a mosaic of CCDs, with each CCD coated with its own custom ARC. Or for single CCD spectrographs, the detector could be coated in sections with an optimized ARC on each section. The section width and resolution of the spectrograph determines the band pass of each region for optimization. As the sections become narrower, or the number of sections on a single CCD is increased, the ARC can approximate a ramp, which would provide a smooth functional form for each layer that varies with wavelength. A ramped coating would be perfectly optimized at each wavelength range, providing a nearly ideal detector for a spectrograph. Such a model, with high QE ($>$ 70\%) across a wide wavelength range is described in Ref.~\citenum{2011Hamden}, although is limited by the changing indices of refraction of silicon.

Using an AR coated, delta-doped CCD in future instrument design can increase throughput and efficiency in several ways. In the case of bright targets, exposure times will be reduced by the increased efficiency of the system, increasing the number of targets that can be explored. Similarly, for the same exposure time or when attempting to detect very low intensity targets, significantly lower detection limits can be reached. When combined with high quality optical coatings designed for high throughput, the effetcs can be considerable- in the case of FIREBall-2, improvements to the detector alone will increase overall system efficiency by more than a factor of 10. An additional gain is in cost savings, especially in the case of space applications. A high efficiency delta-doped detector with a small telescope can do the same or better science at a lower cost than a lower efficiency detector with a large telescope. Reducing the size of the primary mirror, which is a main driver in overall mission costs, will be an important benefit for future mission planning.

Achieving extremely high QE at nearly any UV wavelength is now possible. Coatings of the type described here, with high transmission over a narrow band pass, have many uses. Aside from space applications (astronomy or planetary sciences), other fields such as semiconductor inspection using lasers or medical applications will benefit from the development of these devices. A spectrograph that operates over a narrow range (as in the case of FIREBall), or one with many detectors each observing a narrow range, could also use these types of coatings to achieve overall high QE. As astronomical spectroscopy pushes to dimmer targets (observations of the circumgalactic medium of nearby galaxies, or taking spectra from the atmospheres of extrasolar planets), the high QE from a delta-doped, anti-reflection coated detector will become an essential part of future instruments. 

\appendix

\section{Optical Constants of delta-doped silicon} \label{App:AppendixA}
Here we present measured optical constants for delta-doped silicon. These were measured by J.A. Woollam and are shown compared to normal silicon in Figure \ref{fig:nk}. Indices are found in Table \ref{tab:nk}.

\acknowledgments
The authors wish to thank the reviewers for their detailed, helpful comments and suggestions. In addition, we wish to thank Frank Greer, Michael Lee, and Layton Baker, all of JPL, for their assistance with ALD processes, and Leslie Wulff for helpful discussions. The research was carried out in part at the Jet Propulsion Laboratory, California Institute of Technology, under a contract with NASA. This work was partially supported by KISS, the W. M. Keck Institute for Space Studies, and by NASA Headquarters under the NASA Earth and Space Science Fellowship Program, NASA Grant NNX11AO07H, and NASA Grant NNX12AF29G. Dr. Erika T. Hamden is supported by an NSF Astronomy and Astrophysics Postdoctoral Fellowship under award AST-1402206.

\bibliography{arlib.bib}
\bibliographystyle{spiejour}

\section{Biographies}
Erika~T.~Hamden: Dr. Hamden is a NSF Astronomy and Astrophysics Postdoctoral Fellow, the Caltech Prize Fellow in experimental astrophysics, and a Roman Technology Fellow at Caltech. She is an astrophysicist with degrees from Columbia University (Ph.D., 2014) and Harvard College (A.B., 2006). Her work focuses on UV detector development and instrumentation, with a focus on detecting emission from the diffuse CGM of star forming galaxies. She is the project scientist for the FIREBall-2 mission.

April~D.~Jewell: Dr. Jewell is a technologist at the Jet Propulsion Laboratory. She is a surface scientist with degrees from Tufts University (Ph. D., 2012) and George Washington University (B.S., 2002).  Dr. Jewell’s research interests include the design, development and implementation of device integrated optical coatings, as well as surface passivation techniques, for silicon-based imagers. She holds a Guinness world record for her contribution to discovering the smallest electric motor. 

Charles~A.~Shapiro: Dr. Shaprio is a technologist at the Jet Propulsion Laboratory.

Samuel~R.~Cheng: Mr. Cheng is a technologist at the Jet Propulsion Laboratory

Tim~M.~Goodsall: Dr. Goodsall is a technologist at the Jet Propulsion Laboratory.

John~Hennessy: John Hennessy is a technologist at the Jet Propulsion Laboratory. He received his BE and PhD degrees in electrical engineering from The Cooper Union in 2002, and from the Massachusetts Institute of Technology in 2010, respectively. His current research interests include the development of ALD processes for optical and electrical applications related to UV detector-integrated filters, UV reflective coatings, and semiconductor surface passivation.

Michael~Hoenk: Dr. Michael Hoenk is a Principal Microdevices Engineer in the JPL’s Flight Instrument Detectors group, with extensive experience as principal investigator and leader of JPL research projects on imaging detectors, sensors, and instruments. His many inventions include delta-doped CCDs and superlattice-doped CMOS imaging detectors, which recently resulted in a breakthrough for semiconductor metrology systems. He has received many awards for his work, including the Lew Allen Award for Excellence and the NASA Exceptional Achievement Medal.

Todd~Jones: Dr. Jones is a technologist at the Jet Propulsion Laboratory.

Sam~Gordon: Mr. Gordon is a graduate student at Arizona State University.

Hwei~Ru~Ong: Mr. Ong is a lab technician at Columbia University.

David~Schiminovich: Prof. Schiminovich is a Professor of Astrophysics at Columbia University.

D.~Christopher~Martin: D. Christopher Martin is a Professor of Astrophysics at the California Institute of Technology. He has been building astrophysics experiments for 37 years. He was the PI of NASA’s Galaxy Evolution Explorer, and is currently PI of the Keck Cosmic Web Imager and the FIREBALL-2 Balloon mission.

Shouleh~Nikzad: Shouleh Nikzad is a Senior Research Scientist, a Principal Member of Engineering Staff and the lead for the Advanced Detectors, Systems, and Nanoscience Group at NASA’s Jet Propulsion Laboratory, California Institute of Technology. She has initiated and led the development of detectors, coatings, and devices including current efforts on single photon counting detectors, reflective coatings, and filters. She is a recipient of many awards including a recent NASA Award for development of high efficiency detectors.

\begin{table}[b]
\caption{Table summarizing characteristics of potential high QE coating models. Models 5A and 11A are shown in Figure \ref{fig:11layer}. Other models are shown in Figure \ref{fig:altlayers}.}
\label{tab:char}
\begin{center}
\begin{tabular}{|c|c|c|c|c|c|}
\hline
\rule[-1ex]{0pt}{3.5ex} Model & Materials & \# of  & Max QE & $\lambda$ of Max & Width \\
\rule[-1ex]{0pt}{3.5ex} & & layers & \% & (nm) & at 50\% \\
\rule[-1ex]{0pt}{3.5ex} & & & & & QE \\
\hline
\rule[-1ex]{0pt}{3.5ex} 3A & SiO$_2$,MgF$_2$ & 3 & 61.2 & 204 & 33 \\
\rule[-1ex]{0pt}{3.5ex} 3B & MgO,Al$_2$O$_3$ & 3 & 75.9 & 205 & 37 \\
\rule[-1ex]{0pt}{3.5ex} 3C & SiO$_2$,Al$_2$O$_3$ & 3 & 80.9 & 205 & 47 \\
\rule[-1ex]{0pt}{3.5ex} 3D & MgF$_2$,Al$_2$O$_3$ & 3 & 82.8 & 205 & 48 \\
\rule[-1ex]{0pt}{3.5ex} 3E & MgO,MgF$_2$ & 3 & 87.2 & 205 & 39 \\
\rule[-1ex]{0pt}{3.5ex} 3F & SiO$_2$,MgO & 3 & 87.6 & 205 & 40 \\
\hline
\rule[-1ex]{0pt}{3.5ex} 5A & SiO$_2$,Al$_2$O$_3$ & 5 & 90.4 & 205 & 23 \\
\rule[-1ex]{0pt}{3.5ex} 11A & SiO$_2$,Al$_2$O$_3$ & 11 & 79.7/84.0 & 199/210 & 22 \\
\hline
\end{tabular}
\end{center}
\end{table}

\begin{table}
\caption{Optical Constants of Delta Doped Silicon}
\label{tab:nk}
\begin{center}
\begin{tabular}{|c|c|c||c|c|c||c|c|c||c|c|c|}
\hline
$\lambda$ & n & k & $\lambda$ & n & k & $\lambda$ & n & k & $\lambda$ & n & k \\
(nm) & & & (nm) & & & (nm) & & & (nm) & & \\
\hline
132.6&0.47&2.03&169.9&0.66&2.25&236.2&1.68&3.42&387.5&6.25&0.65\\
133.3&0.61&1.55&171.0&0.67&2.27&238.5&1.69&3.43&393.7&5.96&0.45\\
134.1&0.50&1.73&172.2&0.68&2.30&240.8&1.69&3.45&400.0&5.71&0.33\\
134.8&0.43&1.69&173.4&0.69&2.32&243.1&1.70&3.49&406.6&5.50&0.26\\
135.5&0.49&1.74&174.6&0.70&2.35&245.5&1.70&3.54&413.3&5.32&0.20\\
136.3&0.49&1.73&175.9&0.71&2.37&248.0&1.71&3.61&420.3&5.16&0.16\\
137.0&0.51&1.75&177.1&0.72&2.40&250.5&1.73&3.69&427.6&5.02&0.13\\
137.8&0.51&1.75&178.4&0.74&2.43&253.1&1.75&3.78&435.1&4.91&0.11\\
138.5&0.51&1.79&179.7&0.75&2.45&255.7&1.78&3.87&442.9&4.80&0.09\\
139.3&0.50&1.77&181.0&0.76&2.48&258.3&1.82&3.98&450.9&4.70&0.08\\
140.1&0.52&1.78&182.4&0.78&2.50&261.1&1.87&4.10&459.3&4.61&0.07\\
140.9&0.52&1.78&183.7&0.80&2.53&263.8&1.94&4.24&467.9&4.53&0.06\\
141.7&0.53&1.81&185.1&0.81&2.56&266.7&2.03&4.40&476.9&4.46&0.05\\
142.5&0.54&1.82&186.5&0.83&2.58&269.6&2.14&4.57&486.3&4.39&0.04\\
143.4&0.53&1.84&187.9&0.85&2.61&272.5&2.31&4.76&496.0&4.33&0.04\\
144.2&0.54&1.84&189.3&0.86&2.63&275.6&2.55&4.93&506.1&4.27&0.03\\
145.0&0.55&1.85&190.8&0.88&2.66&278.7&2.82&5.07&516.7&4.22&0.03\\
145.9&0.55&1.87&192.2&0.90&2.68&281.8&3.13&5.18&527.7&4.17&0.03\\
146.7&0.55&1.88&193.8&0.92&2.71&285.1&3.50&5.25&539.1&4.12&0.02\\
147.6&0.56&1.89&195.3&0.93&2.74&288.4&3.94&5.26&551.1&4.08&0.02\\
148.5&0.56&1.91&196.8&0.95&2.77&291.8&4.42&5.09&563.6&4.04&0.02\\
149.4&0.57&1.92&198.4&0.97&2.80&295.2&4.77&4.77&576.7&4.00&0.02\\
150.3&0.57&1.93&200.0&0.99&2.83&298.8&4.98&4.42&590.5&3.96&0.02\\
151.2&0.58&1.95&201.6&1.01&2.86&302.4&5.08&4.10&604.9&3.93&0.01\\
152.1&0.58&1.96&203.3&1.03&2.89&306.2&5.11&3.85&620.0&3.90&0.01\\
153.1&0.59&1.98&205.0&1.06&2.93&310.0&5.12&3.65&635.9&3.87&0.01\\
154.0&0.60&1.99&206.7&1.08&2.96&313.9&5.13&3.49&652.6&3.84&0.01\\
155.0&0.60&2.01&208.4&1.11&2.99&317.9&5.14&3.35&670.3&3.81&0.01\\
156.0&0.61&2.02&210.2&1.13&3.03&322.1&5.16&3.23&688.9&3.79&0.01\\
157.0&0.62&2.03&212.0&1.16&3.06&326.3&5.18&3.14&708.6&3.76&0.01\\
158.0&0.62&2.04&213.8&1.19&3.10&330.7&5.21&3.06&729.4&3.74&0.01\\
159.0&0.63&2.06&215.7&1.22&3.13&335.1&5.25&2.98&751.5&3.72&0.01\\
160.0&0.63&2.07&217.5&1.25&3.17&339.7&5.30&2.91&775.0&3.69&0.01\\
161.0&0.63&2.08&219.5&1.28&3.21&344.4&5.35&2.86&800.0&3.68&0.01\\
162.1&0.64&2.10&221.4&1.32&3.25&349.3&5.42&2.83&826.7&3.66&0.01\\
163.2&0.64&2.11&223.4&1.36&3.30&354.3&5.54&2.85&855.2&3.64&0.01\\
164.2&0.64&2.13&225.5&1.41&3.34&359.4&5.79&2.89&885.7&3.62&0.01\\
165.3&0.65&2.15&227.5&1.46&3.38&364.7&6.25&2.80&918.5&3.61&0.01\\
166.4&0.65&2.17&229.6&1.53&3.42&370.1&6.73&2.31&953.8&3.59&0.01\\
167.6&0.65&2.20&231.8&1.59&3.43&375.8&6.85&1.57&992.0&3.58&0.01\\
168.7&0.65&2.22&234.0&1.65&3.43&381.5&6.59&0.98 & & &\\
169.9&0.66&2.25&236.2&1.68&3.42&387.5&6.25&0.65 & & &\\

\hline
\end{tabular}
\end{center}
\end{table}

\clearpage


\begin{figure}
\begin{center}
\includegraphics[width=0.95\textwidth]{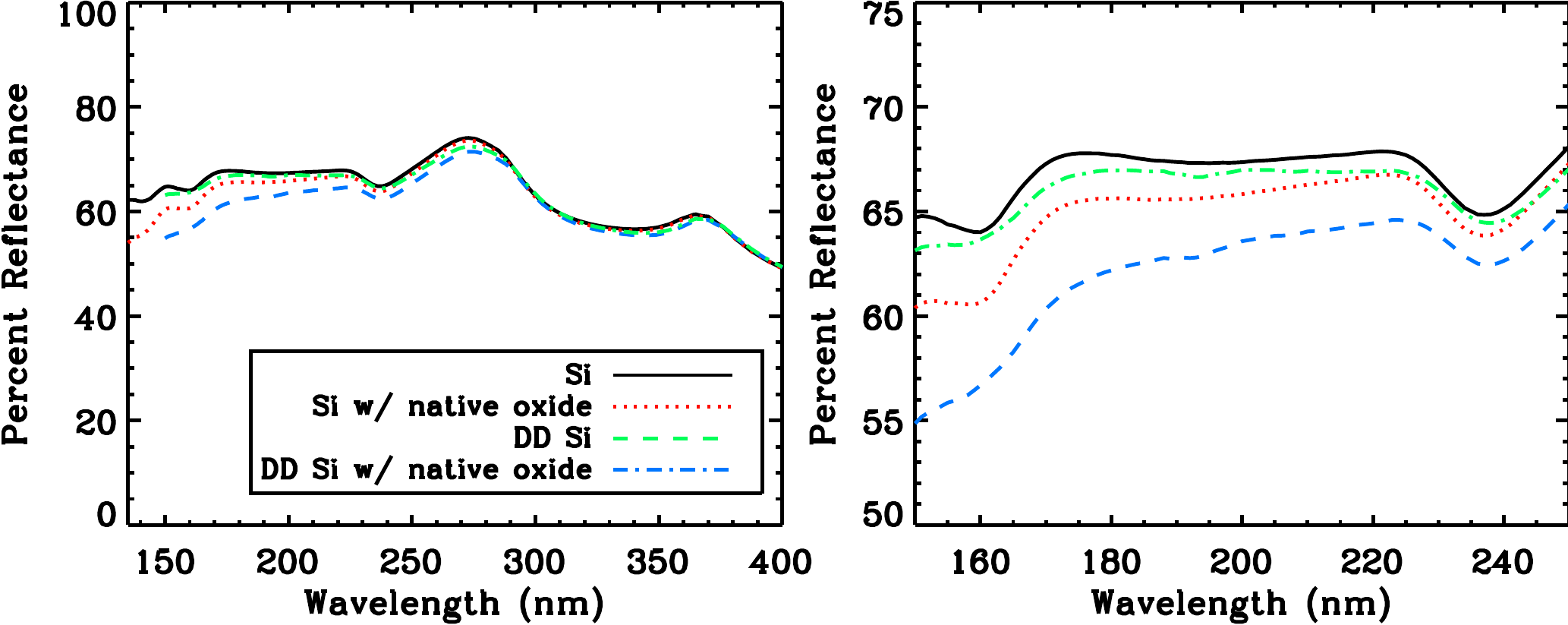}
\caption{\textbf{Left Plot:} Reflectance for Bulk silicon from 137 nm to 400 nm. Over-plotted is the reflectance of silicon with the addition of a 3.413 nm thick delta-doped layer. Models are shown for silicon with a native oxide layer (2 nm SiO$_2$) in both cases. The decreased reflectance below 150 nm for coatings with a native oxide is due completely to absorption in the oxide layer. \textbf{Right Plot:} A close up view of the region of interest for this work, between 140 and 260 nm. In the FIREBall-2 bandpass, the reflectance difference between delta doped and normal silicon is only 0.5\%.}\label{fig:sidd}
\end{center}
\end{figure}

\begin{figure}
\begin{center}
\includegraphics[height=5.5cm]{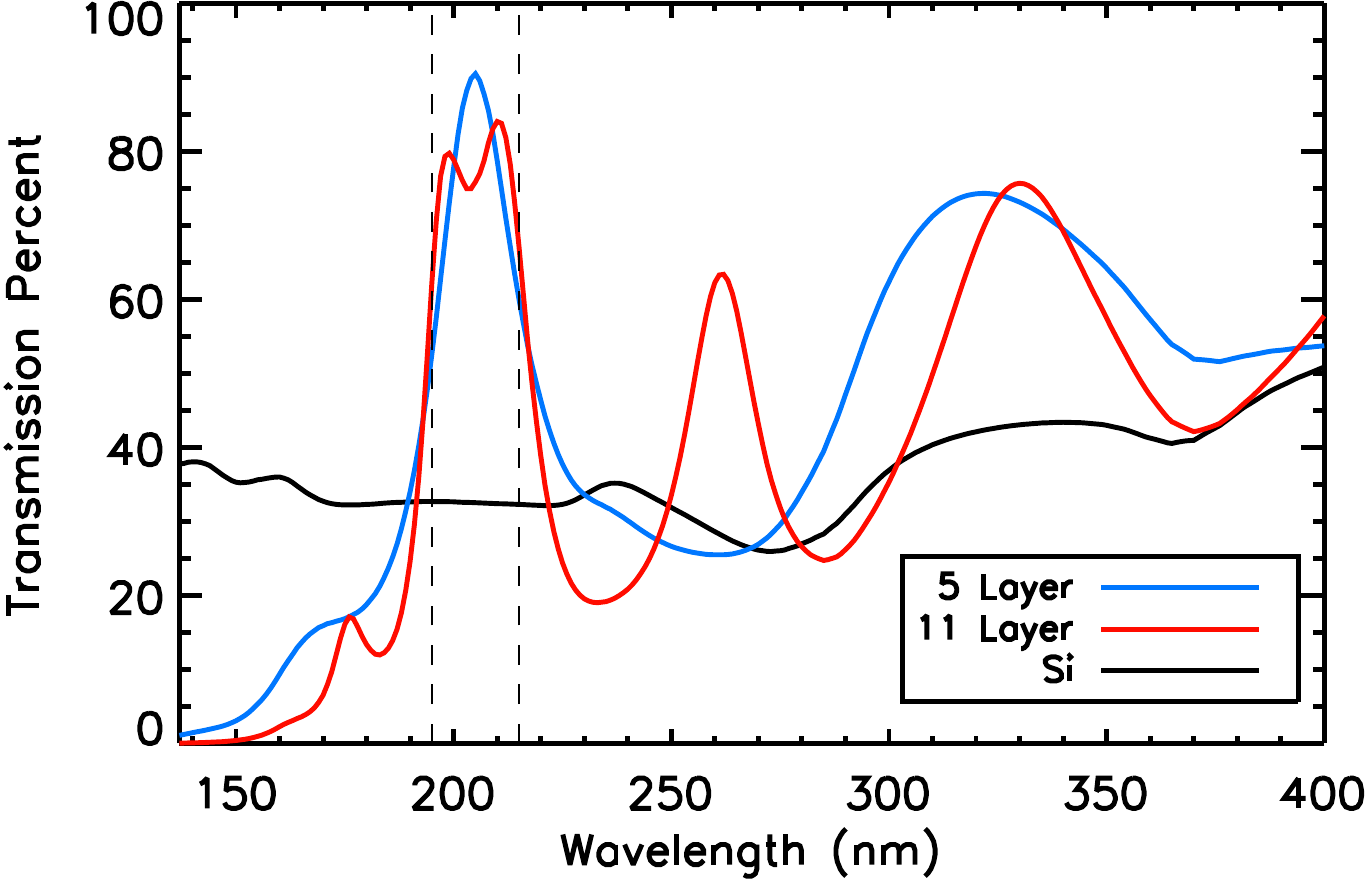}
\caption{Expected transmission vs wavelength for two multi-layers (5A and 11A) optimized around 205 nm. The blue line shows a coating 5A using SiO$_2$ and Al$_2$O$_3$ to achieve a single peak centered around 205 nm. This peak reaches 90.4 \% potential QE. The red line shows coating 11A optimized for high transmission from 195 nm to 215 nm, also using SiO$_2$ and Al$_2$O$_3$. The double peak has a lower maximum, but a higher average transmittance from 195 nm to 215 nm. The black line indicates the reflection limit of bar silicon (100-R).}\label{fig:11layer}
\end{center}
\end{figure}

\begin{figure}
\begin{center}
\includegraphics[height=5.5cm]{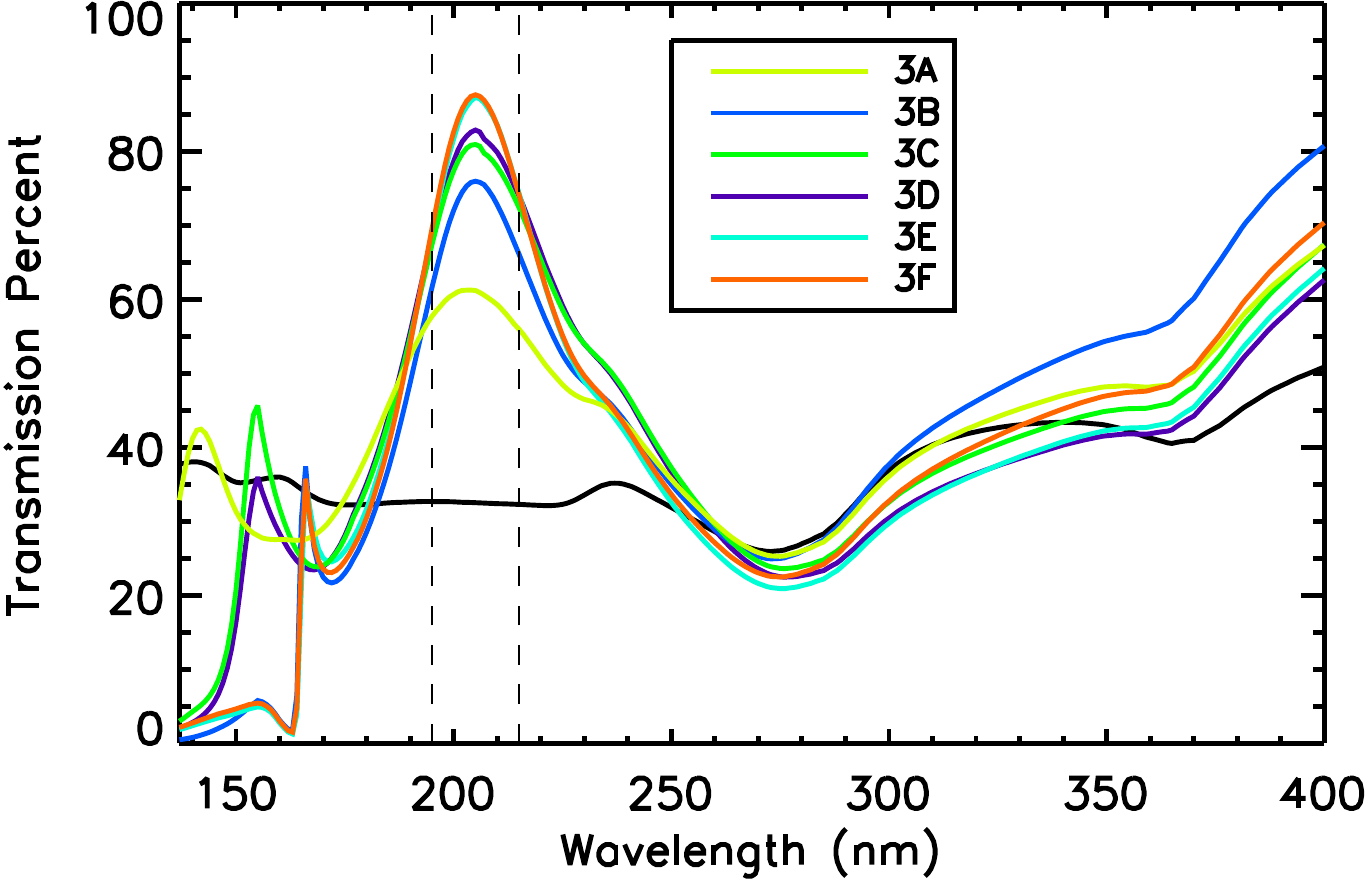}
\caption{Expected transmission vs wavelength for several 3-layer coatings optimized around 205 nm. The reflection limit of silicon is shown in black for reference. Model numbers are described in Table \ref{tab:char}.}\label{fig:altlayers}
\end{center}
\end{figure}

\begin{figure}
\begin{center}
\includegraphics[height=5.5cm]{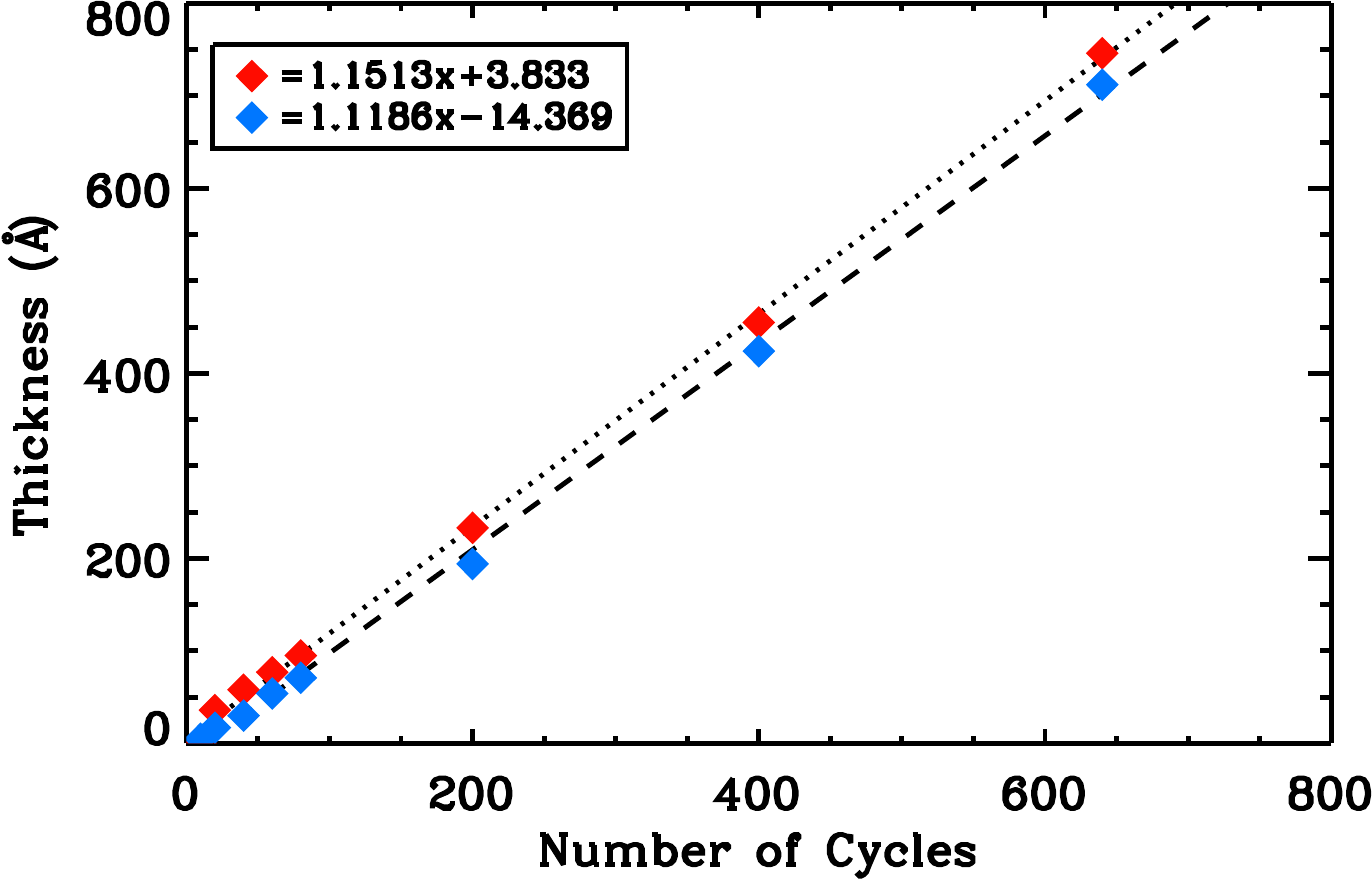}
\caption{Deposition thickness (\AA) of SiO$_2$ on both base silicon (in red) and on a 7 nm thick Al$_2$O$_3$ base layer (in blue) vs. number of ALD cycles. The initial deposition is much slower on the Al$_2$O$_3$ base layer, reflecting the difference in the number of cycles required for good surface nucleation.}\label{fig:nucleation}
\end{center}
\end{figure}

\begin{figure}
\begin{center}
\includegraphics[height=5.5cm]{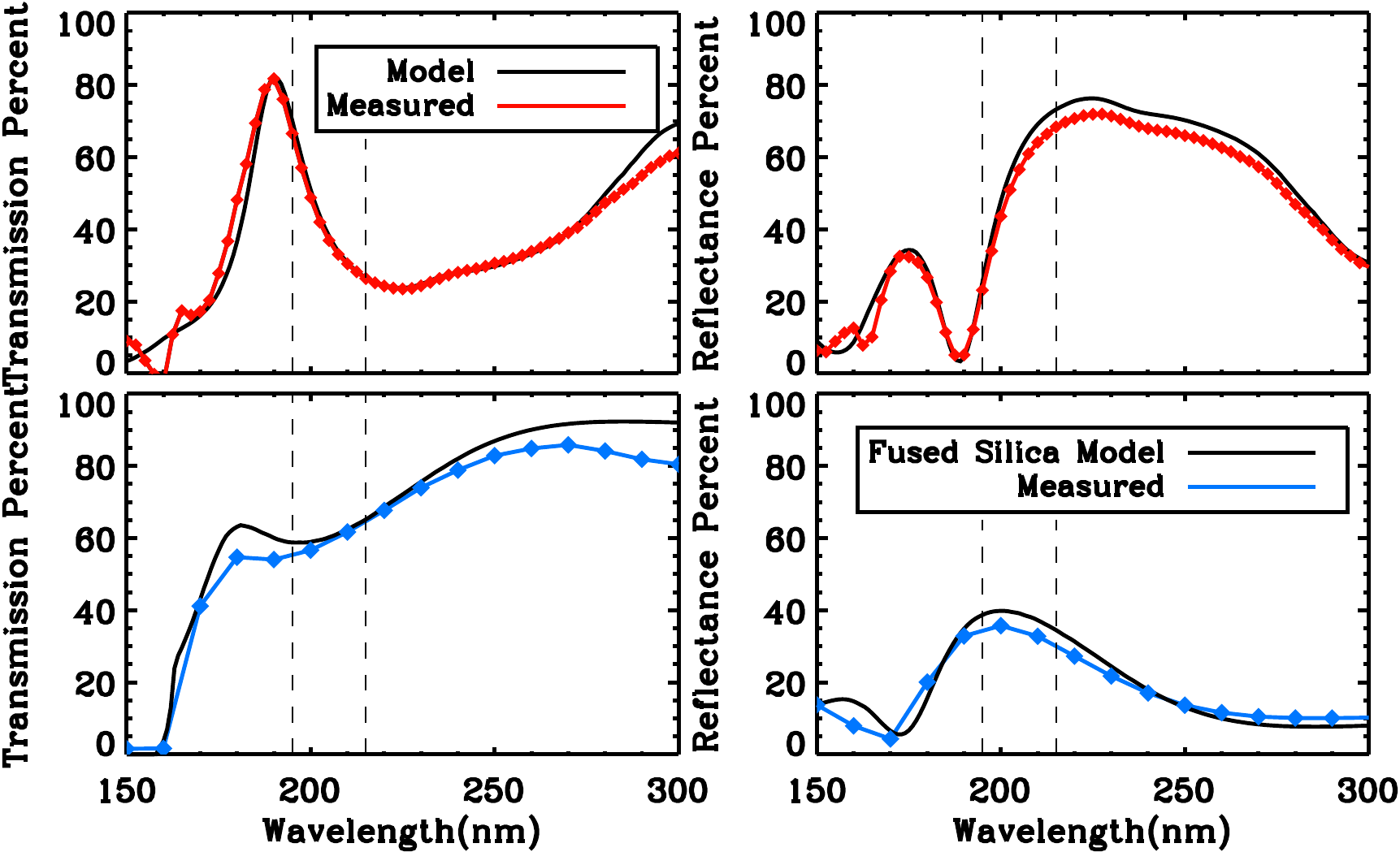} 
\caption{\textbf{Top Left} Calculated transmission for a 5-layer coating (model 5A) of SiO$_2$ and Al$_2$O$_3$. Peak transmission is at 190 nm, and far exceeds previous work at this wavelength (approximately 60\% transmission). This transmission was determined by subtracting the reflectance and absorption from 100 \%, see Equation \ref{eq:transcal1}. \textbf{Top Right} Measured reflectance for a 5-layer coating of SiO$_2$ and Al$_3$O$_3$. Minimum reflectance is at 190 nm. \textbf{Bottom Left} Measured transmission for 5-layer coating on fused silica window. \textbf{Bottom Right} Measured reflectance for 5-layer coating on fused silica window. Both transmission and reflectance of the 11-layer coating on fused silica were used to calculate expected absorption from the coating itself, as described in Section \ref{sec:results}. This absorption measurement is combined with reflectance measurements (Top Right) to calculate expected transmission (Top Left) from the coating when deposited on a silicon device.}\label{fig:5layer}
\end{center}
\end{figure}

\begin{figure}
\begin{center}
\includegraphics[height=5.5cm]{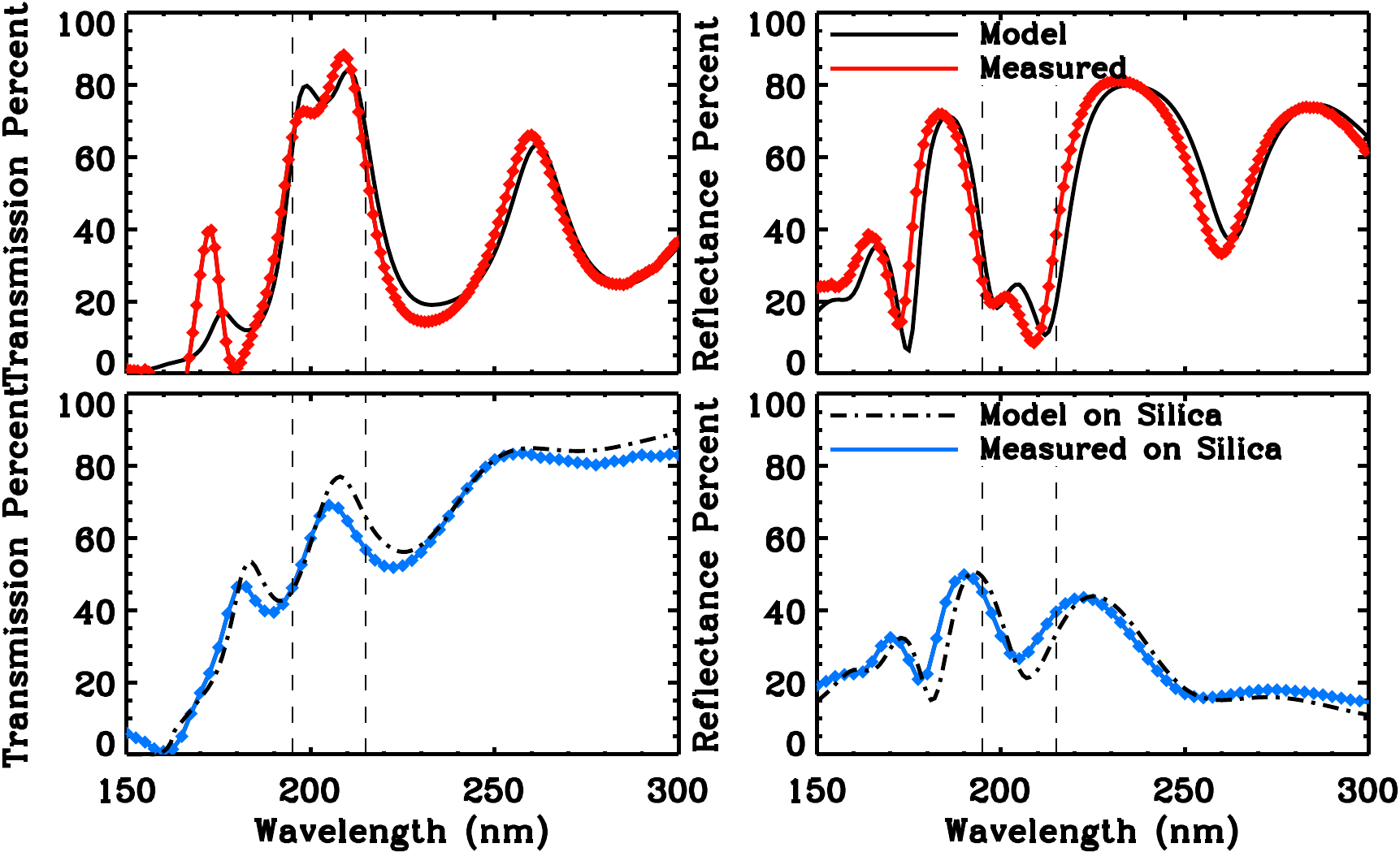}
\caption{\textbf{Top Left} Estimated transmission for an 11-layer coating (model 11A) of SiO$_2$ and Al$_2$O$_3$. Peak transmission is at 209 nm, with an average transmission of 79\% between 195 and 215 nm. This transmission was determined by subtracting the reflectance and absorption from 100 \%. \textbf{Top Right} Measured reflectance for a 11-layer coating of SiO$_2$ and Al$_3$O$_3$. Minimum reflectance is at 209 nm. \textbf{Bottom Left} Measured transmission for 11-layer coating on fused silica window. \textbf{Bottom Right} Measured reflectance for 11-layer coating on fused silica window. Both transmission and reflectance of the 11-layer coating on fused silica were used to calculate expected absorption from the coating itself, as described in Section \ref{sec:results}. This absorption measurement is combined with reflectance measurements (Top Right) to calculate expected transmission (Top Left) from the coating when deposited on a silicon device.}\label{fig:11layer_data}
\end{center}
\end{figure}

\begin{figure}
\begin{center}
\includegraphics[height=5.5cm]{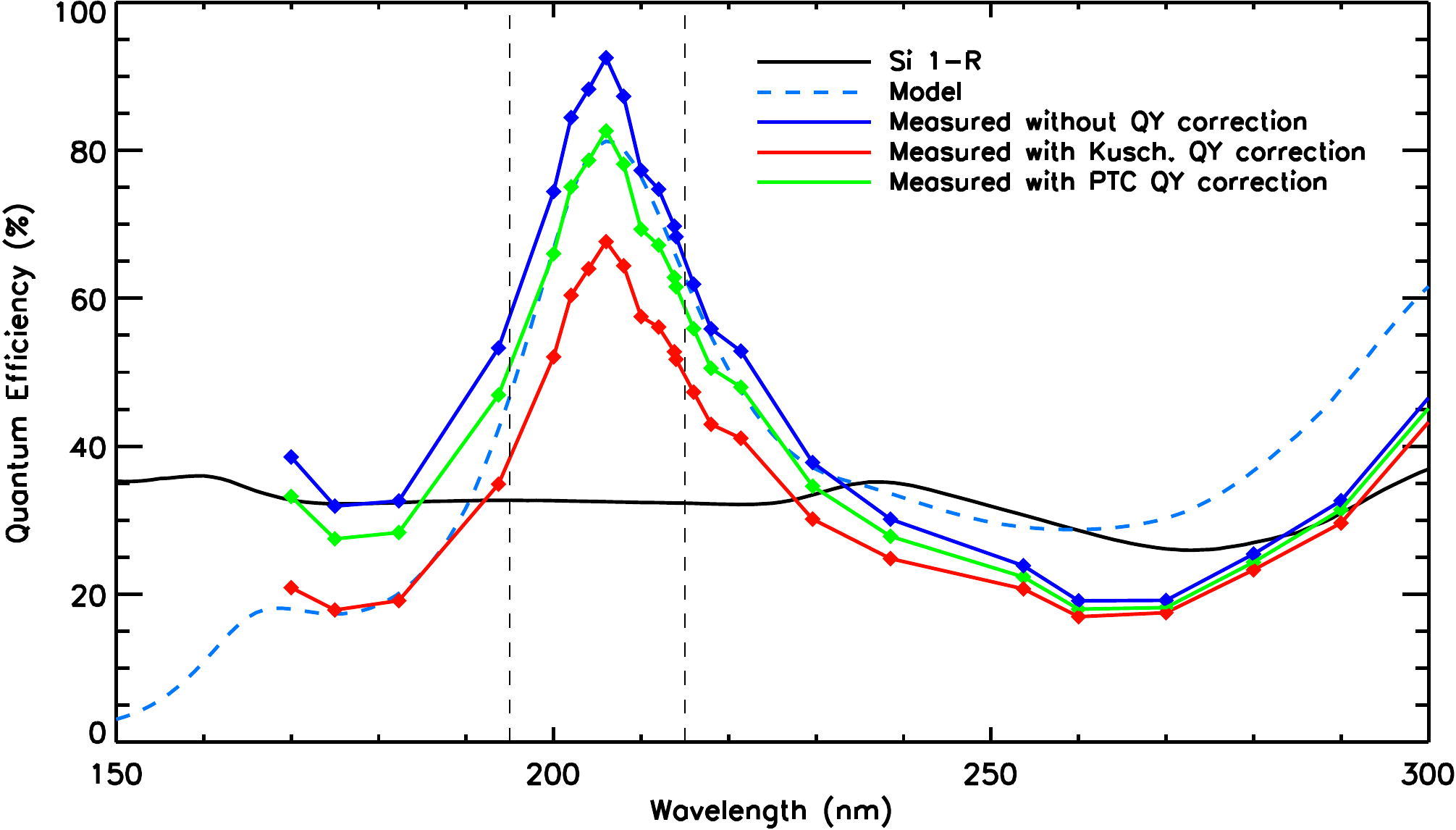}
\caption{Quantum efficiency vs. wavelength for the 5-layer coating deposited on a thinned, delta-doped e2v CCD201-20. The characterization process is described in Section \ref{sec:CCD}. Peak QE is 67.6\% at 206 nm. For comparison, the uncorrected QE measurement is shown in blue. The green line shows the QY corrected QE using the PTC method. This is likely the upper limit of QE values. }\label{fig:QE}
\end{center}
\end{figure}

\begin{figure}
\begin{center}
\includegraphics[height=5.5cm]{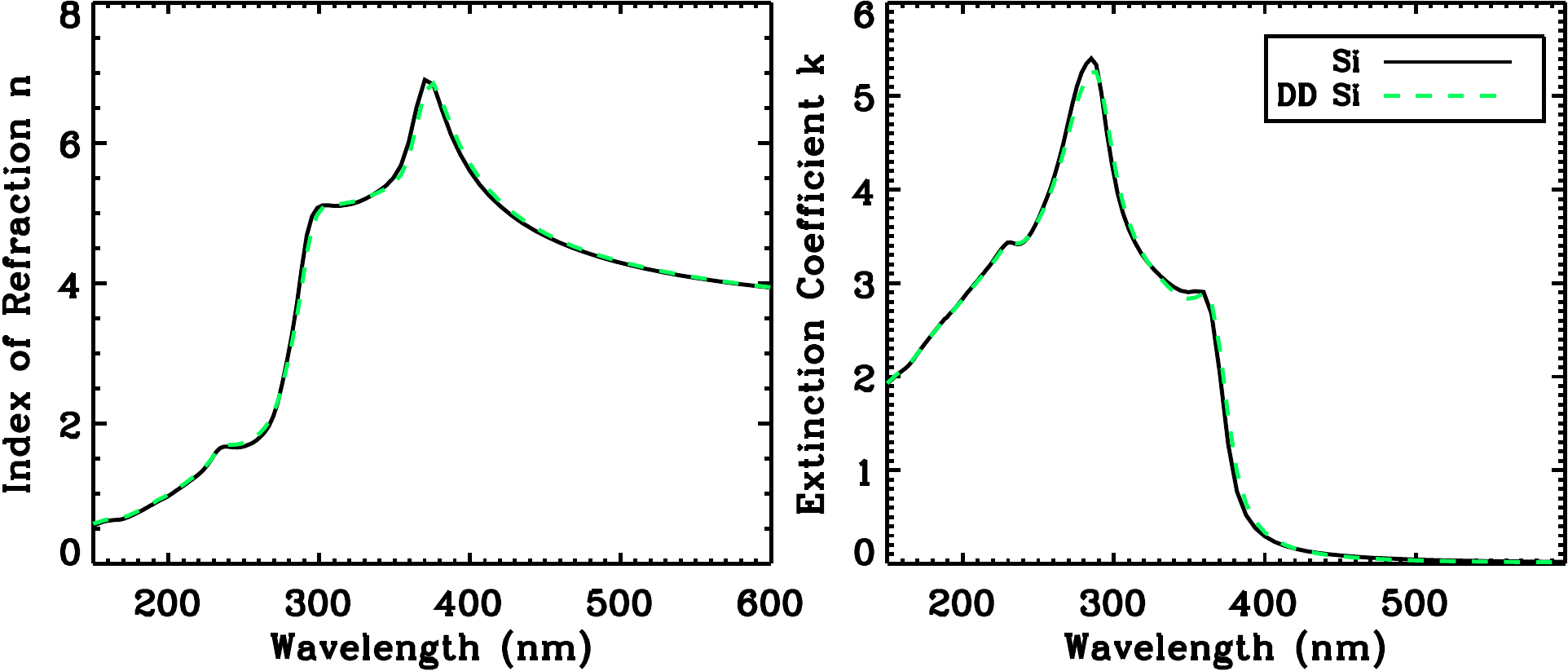}
\caption{Delta doped silicon optical constants are shown in green. Optical constants of normal silicon are shown in black. The differences are slight, but can be important in reflectance calculations.}\label{fig:nk}
\end{center}
\end{figure}

\end{spacing}
\end{document}